\documentclass[11pt,a4paper]{article}
\pdfoutput=1
\usepackage{jheppub}

\usepackage{color}
\usepackage{xcolor}
\usepackage{amsmath}
\usepackage{pifont}   
\usepackage[utf8]{inputenc}

\usepackage{verbatim}   
\usepackage{acronym} 
    
\usepackage{amsfonts}
\usepackage{amssymb}
\usepackage{mathrsfs} 
\usepackage{graphicx}
\usepackage{multirow} 
\usepackage{slashed} 
\usepackage{array}

\usepackage{graphicx}
\usepackage{epsfig}
\usepackage{lscape}
\usepackage{rotating}
\usepackage{epsf}
\usepackage{bm}
\usepackage{booktabs}
\usepackage{array}
\usepackage{caption}
\usepackage{subcaption}
\usepackage[utf8]{inputenc}
\usepackage{float}

\newcommand{\fref}[1]{Fig.~\ref{fig:#1}} 
\newcommand{\eref}[1]{Eq.~\eqref{eq:#1}}

\newcommand{\sref}[1]{Section~\ref{sec:#1}}
\newcommand{\cref}[1]{Chapter~\ref{ch:#1}}
\newcommand{\tref}[1]{Table~\ref{tab:#1}}

\newcommand{\nnl}{\nonumber \\}

\newcommand{\beq}{\begin{equation}} 
\newcommand{\eeq}{\end{equation}} 
\newcommand{\ba}{\begin{array}}  
\newcommand{\ea}{\end{array}} 
\newcommand{\bea}{\begin{eqnarray}}  
\newcommand{\eea}{\end{eqnarray} }  
\newcommand{\be}{\begin{eqnarray}}  
\newcommand{\ee}{\end{eqnarray} }  
\newcommand{\bal}{\begin{align}}
\newcommand{\eal}{\end{align}}   
\newcommand{\bi}{\begin{itemize}}  
\newcommand{\ei}{\end{itemize}}  
\newcommand{\ben}{\begin{enumerate}}  
\newcommand{\een}{\end{enumerate}}  
\newcommand{\bc}{\begin{center}}
\newcommand{\ec}{\end{center}} 
\newcommand{\bt}{\begin{table}}
\newcommand{\et}{\end{table}}  
\newcommand{\btb}{\begin{tabular}}
\newcommand{\etb}{\end{tabular}}

\newcommand{\cO}{{\mathcal O}} 
 
\newcommand{\cL}{{\mathcal L}}

\def\hc{{\rm h.c.}} 

\def\bar#1{\overline{#1}}

\def\inv{^{\raise.15ex\hbox{${\scriptscriptstyle -}$}\kern-.05em 1}}
 
\def\lbar{{\lower.35ex\hbox{$\mathchar'26$}\mkern-10mu\lambda}}

\def\to{\rightarrow}

\let\<=\langle
\let\>=\rangle

\let\+=\uparrow

\begin{document}

\title{Future DUNE constraints on EFT} 
\author[a]{Adam Falkowski,}
\affiliation[a]{Laboratoire de Physique Th\'{e}orique, CNRS, Univ. Paris-Sud, Universit\'{e}  Paris-Saclay, 
\\ 91405 Orsay, France}
\author[b]{Giovanni Grilli di Cortona,}
\author[b]{Zahra Tabrizi}
\affiliation[b]{Instituto de F\'isica, Universidade de S\~ao Paulo, \\C.P. 66.318, 05315-970 S\~ao Paulo, Brazil}
\emailAdd{adam.falkowski@th.u-psud.fr}
\emailAdd{ggrilli@if.usp.br}
\emailAdd{ztabrizi@if.usp.br}

\abstract{
In the near future, fundamental interactions at high-energy scales may be most efficiently studied via precision measurements at low energies.  
A universal language to assemble and interpret precision measurements is the so-called SMEFT, which is an effective field theory (EFT) where the Standard Model (SM) Lagrangian is extended by higher-dimensional operators.  
In this paper we investigate the possible impact of the DUNE neutrino experiment on constraining the SMEFT.  
The unprecedented neutrino flux offers an opportunity to greatly improve the current limits via precision measurements of the trident production and neutrino scattering off electrons and nuclei in the DUNE near detector. 
We quantify the DUNE sensitivity to dimension-6 operators in the SMEFT Lagrangian, and find that in some cases operators suppressed by an $\mathcal{O}(30)$ TeV scale can be probed. 
We also compare the DUNE reach to that of future experiments involving atomic parity violation and polarization asymmetry in electron scattering,  which are  sensitive to an overlapping set of  SMEFT parameters.  
}

\maketitle

\clearpage

\section{Introduction}

Current experimental measurements at high and low  energies are well described  by the  Standard Model (SM) Lagrangian, which respects  the local $SU(3)_C\times SU(2)_L \times U(1)_Y$ symmetry and contains only renormalizable interactions with canonical dimensions $D \leq 4$. 
At the same time, there are several theoretical arguments for the existence of physics beyond the SM (BSM) at some mass scale $\Lambda$. 
It is reasonable to assume that there is a significant gap between the new physics and the weak scales:  
$\Lambda \gg m_Z$. 
If that is the case,  the effects of new particles at the energy scale below $\Lambda$  can be described in a systematic way using  the SM effective field theory (SMEFT).
The framework consists in maintaining the SM particle content and symmetry structure, while abandoning the renormalizability requirement, so that  interaction terms with $D>4$ are allowed \cite{Buchmuller:1985jz}. The higher dimensional operators constructed out of the SM fields describe, in a model independent way,  all possible effects of heavy new physics.
 
In this sense, searching for new physics can be redefined as setting constraints on the Wilson coefficients of higher-dimensional operators in the SMEFT Lagrangian. 
There exist many previous studies along these lines, focusing mostly on the dimension-6 operators (which are expected to give leading contributions for $E \ll \Lambda$). 
A prominent place in this program is taken by precision measurements at very low energies, well below $m_Z$ \cite{Han:2004az,Han:2005pr,Cirigliano:2009wk,Carpentier:2010ue,Filipuzzi:2012mg,Berthier:2015gja,Falkowski:2015krw,Gonzalez-Alonso:2016etj}.
These include neutrino scattering off nucleus or electron targets, atomic parity violation, parity-violating scattering of electrons, and a plethora of meson, nuclear, and tau decay processes.   
For certain dimension-6 operators the low-energy input offers a superior sensitivity compared to that achievable in colliders such as LEP or the LHC, see Ref.~\cite{Falkowski:2017pss} for a recent summary. 
But even in those cases where the new physics reach of low-energy measurements is worse, they often play a vital role in lifting flat directions in the notoriously multi-dimensional parameter space of the SMEFT.  
 
While efforts to get the best out of the existing data continue, it is also important to discuss what progress can be achieved in the future. 
This kind of studies facilitates planning new experiments and analyses, and allow one to understand the complementarity between the low-energy and collider programs.  
In this paper we focus on the future impact of the the Deep Underground Neutrino Experiment (DUNE)  \cite{Acciarri:2015uup}. 
Given the intense neutrino beam, the massive far detector (FD), and the envisaged scale of the near detector (ND), DUNE will certainly offer a rich physics program. 
The main goal is to measure the parameters governing neutrino oscillations: the CP violating phase in the PMNS matrix and the neutrino mass ordering. 
Searches for proton decay and for neutrinos from core-collapse supernovas in our galaxy are also a part of the program. 

From the effective theory perspective, the unprecedented neutrino flux in DUNE offers a unique opportunity to improve the limits on several dimension-6 operators in the SMEFT Lagrangian.  
This can be achieved via precision measurements of various abundant processes in the DUNE ND, such as the production of lepton pairs by a neutrino incident on a target nucleus (the so-called trident production), and neutrino scattering off electrons and nuclei.
In this work we quantitatively study the potential of these processes  to probe the dimension-6 operators. 
Moreover, since the neutrino couplings are related to the charged lepton ones due to the $SU(2)$ local   symmetry of the SMEFT,  we also include in our analysis other future experiments that do not involve neutrinos:  parity-violating M\o ller scattering, atomic parity violation (APV), and polarization asymmetries in scattering of electrons on nuclei.
Finally, we combine our projections with the constraints from existing measurements as described by  the likelihood function obtained  in Ref.~\cite{Falkowski:2017pss}.  
The SMEFT sensitivity of current and future  experiments is compared, with and without the input from DUNE and under different hypotheses about systematic errors.  

The paper is organized as follows. In Section~\ref{sec:Formalism} we explain our formalism. In particular, we review the EFT valid at a few GeV scale relevant for the DUNE experiment,  and we describe its matching with the SMEFT. 
Section~\ref{sec:dune} is devoted to quantitative studies of various neutrino scattering processes in the DUNE ND. Other relevant experimental results which do not involve neutrinos  are discussed in Section~\ref{sec:other}. In Section~\ref{sec:SMEFT} we translate the projected constraints discussed in Sections~\ref{sec:dune} and \ref{sec:other} into the SMEFT language and we give our projections for future constraints. The main results are summarized in Tab.~\ref{tab:dunesmeft} and Fig.~\ref{fig:ressmeft}. 
Section~\ref{sec:conclusions} presents our conclusions.

\section{Formalism}\label{sec:Formalism}

If heavy BSM particles exist in nature, the adequate effective theory at  $E \gtrsim m_Z$ is the so-called SMEFT. 
It has the same particle spectrum and local symmetry as the SM, however the Lagrangian admits higher-dimensional (non-renormalizable) interactions which encode BSM effects. 
However, the SMEFT is not the optimal framework for dealing with observables measured at  $E \ll m_Z$.
At a few GeV scale relevant for DUNE,  the propagating particles are the light SM fermions together with the  $SU(3)_C \times U(1)_{\rm em}$ gauge bosons: the photon and gluons. 
At these energies the $W$, $Z$, and Higgs bosons as well as the top quark can be integrated out,  and their effects can be encoded in contact interactions between the light particles.
To mark  the difference from the SMEFT,  we refer to the effective theory below $m_Z$ as the {\em weak EFT} (wEFT).\footnote{Other names are frequently used in the literature to describe this effective theory, e.g. the {\em Fermi theory} or  the  {\em LEFT} \cite{Jenkins:2017jig}.  } 
It is an effective theory with a limited validity range, and at energies $E \sim m_Z$ it has to be matched to a more complete theory with the full SM spectrum and the larger $SU(3)_C \times SU(2)_L \times U(1)_Y$ local symmetry. 
If the SM were the final theory, that matching would uniquely predict  the wEFT Wilson coefficients in terms of the experimentally well known SM parameters. 
On the other hand, when the wEFT is matched to the SMEFT, the wEFT Wilson coefficients deviate from the SM predictions due to  the effects of the higher-dimensional SMEFT operators. 
This way,   measurements of wEFT parameters in experiments with $E \ll m_Z$ provide non-trivial information about new physics: they allow one to derive constraints on (and possibly to discover) higher-dimensional  SMEFT interactions. 
The latter can be readily translated into constraints on masses and couplings of a large class of BSM theories \cite{deBlas:2017xtg}.  

The future DUNE results are best interpreted in a model-independent way as constraints on the wEFT Wilson coefficients,  but constraints on new physics are more conveniently presented as a likelihood function for the SMEFT Wilson coefficients.
Below we summarize the wEFT interactions relevant for our analysis and the tree-level map between the wEFT and SMEFT coefficients.  

\subsection{Neutrino interactions with charged leptons}

In order to characterize neutrino trident production and neutrino--electron scattering in DUNE we will need the neutrino interactions with electrons and muons. 
In the wEFT, neutrinos interact at tree level with charged leptons via the effective 4-fermion operators:\footnote{
For fermions we use the 2-component spinor notation where a Dirac fermion $F$ is represented as a pair of Weyl spinors $f$, $f^c$: $F = (f, \bar f^c)^T$.
We follow the conventions and notation of Ref.~\cite{Dreiner:2008tw}. 
}
\beq
\label{eq:WEFT_nnll}
\cL_{\rm wEFT} \supset  -  {2 \over v^2}  (\bar \nu_a \bar \sigma_\mu \nu_b) \left [
g_{LL}^{abcd} (\bar e_c \bar \sigma_\mu e_d)
+  g_{LR}^{abcd} (e_c^c \sigma_\mu \bar e_d^c)   \right ],  
\eeq
where the sum over repeated generation indices $a,b,c,d$ is implicit (only the first two generations are relevant for our purpose). 
Integrating out the $W$ and $Z$ bosons at tree level in the SM yields the effective Lagrangian 
\bea
\cL &\supset & 
- { g^{Z \nu}_L g^{Z e}_L  \over m_Z^2} 
(\bar \nu_a \bar \sigma_\mu \nu_a) (\bar e_b \bar \sigma_\mu e_b) 
- { g^{Z \nu}_L g^{Z e}_R  \over m_Z^2} 
 (\bar \nu_a \bar \sigma_\mu \nu_a) (e_b^c \sigma_\mu e_b^c) 
- { (g^{W \ell}_L)^2   \over 2 m_W^2} 
 (\bar \nu_a \bar \sigma_\mu e_a)  (\bar e_b \bar \sigma_\mu \nu_b) 
\nnl & = & 
-  {2 \over v^2}  \left [ 
\left (- {1 \over 2 } + s_\theta^2  \right)  
(\bar \nu_a \bar \sigma_\mu \nu_a) (\bar e_b \bar \sigma_\mu e_b) 
+s_\theta^2   (\bar \nu_a \bar \sigma_\mu \nu_a) (e_b^c \sigma_\mu e_b^c) 
+  (\bar \nu_a \bar \sigma_\mu \nu_b)  (\bar e_b \bar \sigma_\mu e_a)  \right ].
\nnl 
\eea
Here $g^{W \ell}_L = g_L$, $g^{Z f} = T^3_f - s_\theta^2 Y_f$, $s_\theta^2 = g_Y^2/(g_L^2 + g_Y^2)$, 
and $g_L$, $g_Y$ are the gauge couplings of $SU(2)_L \times U(1)_Y$.
Thus, at tree level, the SM  predictions for the wEFT couplings in \eref{WEFT_nnll} are given by 
\bea
\label{eq:WEFT_gsm}
g_{LL, \rm SM}^{abcd} & =  & \left (- {1 \over 2 } + s_\theta^2  \right)  \delta_{ab} \delta_{cd}
+ \delta_{ad} \delta_{bc}, 
\nnl 
g_{LR, \rm SM}^{abcd} & =  &   s_\theta^2 \delta_{ab} \delta_{cd}. 
\eea 
For the numerical SM values we use  $g_{LL, \rm SM}^{2211} = (g_{LV, \rm SM}^{\nu_\mu e}  + g_{LA, \rm SM}^{\nu_\mu e} )/2=  -0.2730$, $g_{LR, \rm SM}^{2211} = (g_{LV, \rm SM}^{\nu_\mu e}  - g_{LA, \rm SM}^{\nu_\mu e} )/2=0.2334$ \cite{Olive:2016xmw}, which incorporates some loop correction effects.  
For the remaining operators we use the analytic expression in terms of $s_\theta^2$ evaluated at the central value of  the low-energy Weinberg angle $s_\theta^2=0.23865$ \cite{Olive:2016xmw}.  

Going beyond the SM we have $g_{LX}^{abcd} = g_{LX, \rm SM}^{abcd} + \delta g_{LX}^{abcd}$, where $\delta g_{LX}^{abcd}$ can be calculated in terms of some high-energy parameters once the UV completion of the wEFT is specified.\footnote{In the neutrino literature BSM effects in the wEFT Lagrangian are often referred to as non-standard interactions (NSI) and parametrized by  $\epsilon^{ff^\prime X}_{\alpha\beta}$.
The translation between that  language and our formalism is simple: $\epsilon_{ii}^{eX}=\delta g^{ii11}_{LX}$, with  $\epsilon_{ii}^{eX}$ defined as in Ref. \cite{Farzan:2017xzy}.}
Here we assume that the wEFT is matched  at $E \sim m_Z$ to the SMEFT Lagrangian truncated at the level of dimension-6 operators. 
Moreover, for the purpose of studying neutrino scattering at GeV energies, one can safely ignore the dimension-5 SMEFT operators which give masses to the neutrinos. 
Thus we consider the Lagrangian  ${\cal L}~=~{\cal L}_{\rm SM}~+~\sum_{i}  \frac{c_{i}}{v^2} O_{i}^{D=6}$, 
where $ {\cal L}_{\rm SM}$ is the SM Lagrangian, $v=(\sqrt{2}G_F)^{-1/2}\simeq 246$~GeV,  each $O_i^{D=6}$ is a gauge-invariant operator of dimension $D$=6, and $c_i$ are the corresponding Wilson coefficients.
We will work consistently up to linear order in $c_i$, neglecting quadratic and higher powers.  
In full generality, such a framework introduces 2499 new  independent free parameters \cite{Grzadkowski:2010es,Alonso:2013hga}, but working at tree level only a small subset of those is relevant for our analysis. 
The relevant parameter space can be conveniently characterized 
by a set of vertex corrections $\delta g$ to the $Z$ and $W$ interactions with leptons, and by Wilson coefficients of 4-lepton operators \cite{Gupta:2014rxa,deFlorian:2016spz}. 
The former are defined via the Lagrangian 
\bea
\label{eq:SMEFT_dg}
{\cal L}_{\rm SMEFT} & \supset  & 
 {g_L \over \sqrt 2} \left[ 
 W^{\mu+}  \bar \nu_a \bar \sigma_\mu (1 +  \delta g^{W e_a}_L  ) e_a + \hc  \right ]
+ \sqrt{g_L^2 + g_Y^2} Z^\mu  e_a^c \sigma_\mu \left (   - s^2_\theta Q_f  + \delta g^{Z e_a}_R \right ) \bar e_a^c 
 \nnl &+& 
  \sqrt{g_L^2 + g_Y^2} Z^\mu    
\sum_{f = e,\nu} \bar f_a \bar \sigma_\mu \left ( T_3^f -s_\theta^2 Q_f + \delta g^{Z f_a}_L  \right ) f_a , 
\eea
where we display only the flavor-diagonal interactions. 
Not all the vertex corrections above are independent, as in the dimension-6 SMEFT there is the relation $\delta g^{Z\nu_a}_L -  \delta g^{Ze_a}_L  = \delta g^{W e_a}_L$.  
The vertex corrections can be expressed by a combination of dimension-6 Wilson coefficients in any operator basis  (see e.g. \cite{Falkowski:2017pss} for the map to the so-called Warsaw basis), but it is much  more convenient to span the relevant parameter space with $\delta g$'s.  
The remaining parameters we make use here are  Wilson coefficients of the 4-lepton operators  collected in \tref{4l}.  
Note that only the ones containing the left-handed lepton doublet $\ell = (\nu,e)^T$ give rise to neutrino interactions, but for completeness we also list the ones without $\ell$.

\begin{table}
\bc
\begin{tabular}{c|c}
\hline 
One flavor ($a=1,2,3$) & Two flavors ($a < b =1,2,3$) \\ 
\hline 
&
\\
$ [O_{\ell \ell}]_{aaaa} = {1\over 2} (\bar \ell_a\bar \sigma_\mu \ell_a)  (\bar \ell_a \bar \sigma^\mu \ell_a)$
 & $ [O_{\ell \ell}]_{aabb}  =  (\bar \ell_a\bar \sigma_\mu \ell_a)  (\bar \ell_b \bar \sigma^\mu \ell_b) $ 
 \\ 
  & $[O_{\ell \ell}]_{abba} = (\bar \ell_a \bar \sigma_\mu \ell_b)  (\bar \ell_b \bar \sigma^\mu \ell_a)  $
 \\   
$ [O_{\ell e}]_{aaaa} =  (\bar \ell_a\bar \sigma_\mu \ell_a)  (e_a^c  \sigma^\mu \bar e_a^c) $ &
$ [O_{\ell e}]_{aabb}  =  (\bar \ell_a\bar \sigma_\mu \ell_a)  (e_b^c  \sigma^\mu \bar e_b^c)$
  \\
  &
$ [O_{\ell e}]_{bbaa}  =  (\bar \ell_b \bar \sigma_\mu \ell_b)  (e_a^c  \sigma^\mu \bar e_a^c)$
  \\
 &  $[O_{\ell e}]_{abba}  =  (\bar \ell_a \bar \sigma_\mu \ell_b)  (e_b^c  \sigma^\mu \bar e_a^c)$
\\ 
 $ [O_{e e}]_{aaaa} =   {1\over 2} (e_a^c  \sigma_\mu \bar e_a^c)   (e_a^c  \sigma^\mu \bar e_a^c) $ &
 $ [O_{e e}]_{aabb}  =   (e_a^c  \sigma_\mu \bar e_a^c)   (e_b^c  \sigma^\mu \bar e_b^c) $
 \end{tabular}
\ec 
\caption{Flavor-conserving 4-lepton operators in the SMEFT Lagrangian.
\label{tab:4l}
}
\end{table}

We are ready to write down the tree-level matching equations between the wEFT and SMEFT parameters. 
For the wEFT couplings relevant for our analysis we find   
\bea
\label{eq:gll1111}
g_{LL, \rm SM}^{1111} & = &   {1 \over 2} + s_\theta^2, \qquad  g_{LR, \rm SM}^{1111}  = s_\theta^2,  
\nnl 
\delta g_{LL}^{1111}  & = & 
- {1 \over 2} [c_{\ell \ell}]_{1111} +  {1 \over 2} [c_{\ell \ell}]_{1221}  -  \delta g^{W \mu}_L 
 + 2 s_\theta^2 \left ( \delta g^{W e}_L +  \delta g^{Z e}_L  \right ),
\nnl 
\delta g_{LR}^{1111}  & = & 
- {1 \over 2} [c_{\ell e}]_{1111}
+ \delta g^{Z e}_R + 2 s_\theta^2 \left ( \delta g^{W e}_L +  \delta g^{Z e}_L  \right ). 
\eea 
\bea
g_{LL, \rm SM}^{1122} & = &   - {1 \over 2} + s_\theta^2, \qquad  g_{LR, \rm SM}^{1122}  = s_\theta^2,  
\nnl 
\delta g_{LL}^{1122}  & = & 
- {1 \over 2} [c_{\ell \ell}]_{1122}   +  \delta g^{Z \mu}_L 
 + (-1 + 2 s_\theta^2) \left ( \delta g^{W e}_L +  \delta g^{Z e}_L  \right ),
\nnl 
\delta g_{LR}^{1122}  & = & 
- {1 \over 2} [c_{\ell e}]_{1122}
+ \delta g^{Z \mu}_R + 2 s_\theta^2 \left ( \delta g^{W e}_L +  \delta g^{Z e}_L  \right ). 
\eea 
\bea
\label{eq:gll1221}
g_{LL, \rm SM}^{1221} = g_{LL, \rm SM}^{2112}  & = &   1,  \qquad  g_{LR, \rm SM}^{1221} = g_{LR, \rm SM}^{2112}= 0,  
\nnl 
\delta g_{LL}^{1221} = \delta g_{LL}^{2112} & =  & \delta g_{LR}^{1221} = \delta g_{LR}^{2112}  =   0. 
\eea
\bea
g_{LL, \rm SM}^{2211} & = &   - {1 \over 2} + s_\theta^2, \qquad  g_{LR, \rm SM}^{2211}  = s_\theta^2,  
\nnl 
\delta g_{LL}^{2211}  & = & 
- {1 \over 2} [c_{\ell \ell}]_{1122}   +  \delta g^{Z e}_L 
 + (-1 + 2 s_\theta^2) \left ( \delta g^{W \mu}_L +  \delta g^{Z \mu}_L  \right ),
\nnl 
\delta g_{LR}^{2211}  & = & 
- {1 \over 2} [c_{\ell e}]_{2211}
+ \delta g^{Z e}_R + 2 s_\theta^2 \left ( \delta g^{W \mu}_L +  \delta g^{Z \mu}_L  \right ). 
\eea 
\bea
\label{eq:gll2222}
g_{LL, \rm SM}^{2222} & = &   {1 \over 2} + s_\theta^2, \qquad  g_{LR, \rm SM}^{2222}  = s_\theta^2,  
\nnl 
\delta g_{LL}^{2222}  & = & 
- {1 \over 2} [c_{\ell \ell}]_{2222} +  {1 \over 2} [c_{\ell \ell}]_{1221}  -  \delta g^{W e}_L 
 + 2 s_\theta^2 \left ( \delta g^{W \mu}_L +  \delta g^{Z \mu}_L  \right ),
\nnl 
\delta g_{LR}^{2222}  & = & 
- {1 \over 2} [c_{\ell e}]_{2222}
+ \delta g^{Z \mu}_R + 2 s_\theta^2 \left ( \delta g^{W \mu}_L +  \delta g^{Z \mu}_L  \right ). 
\eea 
A part of the results above may appear counterintuitive. 
For example $\delta g_{LL}^{1111}$ depends on  $[c_{\ell \ell}]_{1221}$ and $\delta g^{W \mu}_L$, 
and likewise $\delta g_{LL}^{2222} $ depends on  $\delta g^{W e}_L$. 
This happens because some dimension-6 SMEFT operators  affect the observables $G_F$, $\alpha(0)$, and $m_Z$ which traditionally serve as the input to determine the SM parameters $g_L$, $g_Y$ and $v$ from experiment.  
To take this into account one needs to absorb this effect into a redefinition of the SM parameters, which brings new terms into the matching equation. 
The most dramatic consequence is that  $\delta g_{LL}^{1122}$ and  $\delta g_{LL}^{2211}$ do not depend on new physics at all. 
That is because the corresponding 4-fermion  SMEFT operator is responsible for the muon decay, from which  the Fermi constant $G_F$ is experimentally determined.

\subsection{Neutrino interactions with quarks}

Another class of processes relevant in DUNE is the charged current (CC) and neutral current (NC) scattering of neutrinos on atomic nuclei.
This can be characterized by 4-fermions wEFT  interactions of  neutrinos with up and down quarks: 
\beq 
\label{eq:WEFT_nnqq}
\cL_{\rm wEFT} \supset -  {2 \tilde{V}_{ud} \over v^2} (1+\bar \epsilon_L^{d e_a})  (\bar e_a \bar \sigma_\mu \nu_a) ({\bar u}\, {\bar \sigma}^\mu d)
- {2  \over v^2} (\bar \nu_a \bar \sigma_\mu \nu_a)  \sum_{q=u,d} 
\left [
g_{LL}^{\nu_a q} \,\bar q\, \bar \sigma^\mu q
+  g_{LR}^{\nu_a q} (q^c \sigma^\mu \bar q^c)   \right ]. 
\eeq
We do not display here the CC interactions of right-handed quark and  chirality-violating NC interactions, as their effects for neutrino scattering in DUNE are suppressed.  
$\tilde V_{ud}$ denotes the  CKM matrix defined in  such  a way that most of new physics corrections affecting observables from which $V_{ud}$ is experimentally determined are absorbed in its definition \cite{Gonzalez-Alonso:2016etj}. 
In the following we will use the numerical value $\tilde V_{ud} =0.97451$, which is the central value of the fit in  \cite{Gonzalez-Alonso:2016etj}, although more generally in a global analysis one should fit $\tilde V_{ud}$ simultaneously with new physics parameters.  
In the SM the wEFT parameters in \eref{WEFT_nnqq} take the values 
$g_{LL,SM}^{\nu_a u} = \frac{1}{2}-\frac{2 s_\theta^2}{3}$, 
$g_{LR,SM}^{\nu_a u} = -\frac{2 s_\theta^2}{3}$,
$g_{LL,SM}^{\nu_a d} = -\frac{1}{2}+\frac{s_\theta^2}{3}$,  
$g_{LR,SM}^{\nu_a d} = \frac{s_\theta^2}{3}$, 
and $ \bar \epsilon_{L, \rm SM}^{d e_a} = 0$. 
Neutrino scattering experiments typically measure ratios of NC to CC neutrino or anti-neutrino cross sections. For this reason, it is  convenient to introduce the following notation: 
\beq
\label{eq:gnuadef}
\left( g_{L/R}^{\nu_a} \right)^2 \equiv \frac{\left( g_{LL/LR}^{\nu_a u} \right)^2 + \left( g_{LL/LR}^{\nu_a d} \right)^2}{\left( 1+  \bar \epsilon_L^{d e_a}  \right)^2}, \qquad 
\tan \theta_{L/R}^{\nu_a} \equiv \frac{g_{LL/LR}^{\nu_a u}}{g_{LL/LR}^{\nu_a d}}, 
\eeq
where the SM predictions are 
$(g_{L, \rm SM}^{\nu})^2 = 0.3034$, 
$(g_{R, \rm SM}^{\nu})^2 = 0.0302$,  
$\tan \theta_{L, \rm SM}^{\nu} = -0.80617$, 
$\tan \theta_{R, \rm SM}^{\nu} = -1.9977$  \cite{Olive:2016xmw}.

\begin{table}
\bc
\begin{tabular}{c|c}
\hline 
With lepton doublets & Without lepton doublets   \\ 
\hline 
\\
$ [O_{\ell q}]_{aabb} = (\bar \ell_a\bar \sigma_\mu \ell_a)  (\bar q_b \bar \sigma^\mu q_b)$
 & $ [O_{e q}]_{aabb} = (e^c_a \sigma_\mu \bar e^c_a)  (\bar q_b \bar \sigma^\mu q_b)$ 
 \\ 
$ [O^{(3)}_{\ell q}]_{aabb} = (\bar \ell_a \bar \sigma_\mu \sigma^i \ell_a)  (\bar q_b \bar \sigma^\mu \sigma^i q_b)$ 
& $ [O_{e u}]_{aabb} = (e^c_a \sigma_\mu \bar e^c_a)  (u^c_b \sigma^\mu \bar u^c_b)$ 
 \\   
$ [O_{\ell u}]_{aabb} = (\bar \ell_a \bar \sigma_\mu \ell_a)  (u^c_b \sigma^\mu \bar u^c_b)$ 
 & $ [O_{e d}]_{aabb} = (e^c_a \sigma_\mu \bar e^c_a)  (d^c_b \sigma^\mu \bar d^c_b)$ 
 \\
$ [O_{\ell d}]_{aabb} = (\bar \ell_a \bar \sigma_\mu \ell_a)  (d^c_b \sigma^\mu \bar d^c_b)$ &
  \\
 \end{tabular}
\ec 
\caption{Flavor- and chirality-conserving 2-lepton-2-quark operators in the SMEFT Lagrangian.
\label{tab:2l2q}
}
\end{table}

Turning to the SMEFT, much as in the 4-lepton case before, the space of relevant dimension-6 operators can be parameterized by vertex corrections and a set of 4-fermion operators. 
The former are defined by 
\bea
\label{eq:SMEFT_dgq}
{\cal L}_{\rm SMEFT} & \supset  & 
\sqrt{g_L^2 + g_Y^2} Z^\mu  \sum_{q=u,d} \big [  
\bar q \bar \sigma_\mu \left (  (T_3^q -s_\theta^2 Q_q) + \delta g^{Zq}_L \right ) q  
+  q^c \sigma_\mu \left (   - s^2_\theta Q_q + \delta g^{Z q}_R\right ) \bar q^c  \big ]
\nnl &+& 
\left [ W^{\mu+}  \bar u \bar \sigma_\mu \left(V_{ud} +   \delta g^{W q_1}_L  \right) d +\hc \right ]  .
\eea
The relevant $LLQQ$ 4-fermion operators are summarized in \tref{2l2q}. 
Only the ones containing lepton doublets are relevant for neutrino interactions, but for future references we have also displayed the operators affecting only charged lepton couplings to quarks. 
Writing $g_{LX}^{\nu_a q} =  g_{LX,SM}^{\nu_a q}  + \delta g_{LX}^{\nu_a q}$ and matching at tree level to the SMEFT Lagrangian at $E\sim m_Z$ we obtain the following relations between the parameters of the two theories:
\bea
\delta g_{LL}^{\nu_a u} &=& \delta g_L^{Zu} + \left(1-\frac{4 s_\theta^2}{3} \right) \delta g_L^{Z \nu_a} - \frac{1}{2} ([c_{\ell q}]_{aa11} + [c_{\ell q}^{(3)}]_{aa11}),
\nnl
\delta g_{LR}^{\nu_a u} &=&  \delta g_R^{Zu} -\frac{4 s_\theta^2}{3}  \delta g_L^{Z \nu_a} - \frac{1}{2} [c_{\ell u}^{(3)}]_{aa11},
\eea
\bea
\delta g_{LL}^{\nu_a d} &=& \delta g_L^{Zd} - \left(1-\frac{2 s_\theta^2}{3} \right) \delta g_L^{Z \nu_a} - \frac{1}{2} ([c_{\ell q}]_{aa11} - [c_{\ell q}^{(3)}]_{aa11}),
\nnl
\delta g_{LR}^{\nu_a d} &=&  \delta g_R^{Zd} + \frac{2 s_\theta^2}{3}  \delta g_L^{Z \nu_a} - \frac{1}{2} [c_{\ell d}^{(3)}]_{aa11}, 
\eea
\bea
 \bar \epsilon_L^{d e} &= & -  \delta g_R^{W q_1} , 
\nnl 
 \bar \epsilon_L^{d \mu}   &= &  
 -  \delta g_R^{W q_1}  +  \delta g_L^{W \mu}  -  \delta g_L^{W e} + [c^{(3)}_{lq}]_{111 1} -  [c^{(3)}_{lq}]_{2211}.  
 \eea 
 Above, we work in the approximation where the off-diagonal CKM parameters are treated as zero when they multiply dimension-6 SMEFT parameters; see   \cite{Falkowski:2017pss} for more general expressions. 
Again, a somewhat counterintuitive expressions for  $\bar \epsilon_L^{d e}$ follows from absorbing part new physics effects into the definition $\tilde V_{ud}$.
In the neutrino literature the BSM effects in the wEFT Lagrangian are often referred to as non-standard interactions (NSI) and parametrized by  $\epsilon^{ff^\prime X}_{\alpha\beta}$.
The translation to the NSI language is  $\epsilon^{udL}_{aa}=\tilde V_{ud}\bar \epsilon_L^{d e_a}$ and $\epsilon^{qX}_{aa}=\delta g^{\nu_aq}_{LX}$,  with  $\epsilon_{aa}$'s defined as in Ref. \cite{Farzan:2017xzy}.

\section{Neutrino scattering in DUNE}
\label{sec:dune} 

The DUNE design includes a near detector (ND)  located at $574$ m from the proton beam target, as well as a far detector (FD) with $34$ kilotonnes of argon mass  at $1297$ km from the source.
The ND is primarily designed to provide constraints on the systematic uncertainties in oscillation studies. 
On the other hand, thanks to the  extremely high rate of neutrino interactions, it can be readily used for precision cross section measurements.

For the following analysis we assume $3$+$3$ years of operation for neutrino+antineutrino beams  and a near detector of 100 tonnes argon mass. 
Each beam of neutrinos (antineutrinos) consists of approximately $90\%$ $\nu_\mu~(\bar{\nu}_\mu)$ beams and $10\%$ contamination of antineutrinos (neutrinos). 
In addition, each beam contains less than $1\%$ of $\nu_e$ and $\bar{\nu}_e$ from pion decays.  
Our analysis is based on calculating the number of events at DUNE, for which we use the neutrino fluxes given in Ref. \cite{Alion:2016uaj}. 
The neutrino energies in DUNE range from $0.25$ GeV to $20$ GeV, 
however for our analysis we consider  $0.25$~GeV bins between $0.25$ GeV and $8.25$ GeV  as the contribution from the  higher energies to the total flux is negligible.    
We calculate the expected number of events as:
\begin{eqnarray}
\label{eq:nevents}
N={\rm{time}}\times{\#\rm{~of~targets}}\times {\rm{efficiency}}\times\int_{E_i}^{E_f}dE_\nu \frac{d\phi(E_\nu)}{dE_\nu}\sigma(E_\nu)\,,
\end{eqnarray}
where $\phi$ is the neutrino flux, $\sigma$ is the relevant cross section, $E_\nu$ is the neutrino energy,  $E_i = 0.25$~GeV and $E_f = 8.25$ GeV. Here the number of targets is calculated for $1.1 \times 10^{21}$ POT (proton on target) in (anti-)neutrino mode with a 120-GeV proton beam with 1.20 MW of power. 

In the following we describe the relevant observables at DUNE for  trident production, elastic neutrino scattering on electrons and neutrino scattering off nuclei.

\subsection{Trident production}

We start with the {\em trident} production in DUNE ND. 
The trident events are the production of lepton pairs by a neutrino incident on a heavy nucleus: 
$\nu_a  N \to \nu_b e_c^+ e_d^- N$, 
and their potential to probe new physics was emphasized  in  Ref.~\cite{Altmannshofer:2014cfa,Altmannshofer:2014pba}.  
The trident process has been previously observed by the CHARM-II \cite{Vilain:1994qy} and CCFR \cite{Mishra:1991bv} experiments.  However, due to technological limitations in the detector design, these experiments could access the $\nu \mu^+ \mu^-$ final state and the accuracy of the cross section measurement was only $\cO(50)$\%.

The cross sections in the SM for different neutrino channels are taken from \cite{Magill:2016hgc}. 
We assume flat neutrino efficiencies, $85\%$ for $\nu_\mu~(\bar{\nu}_\mu)$ and $80\%$ for $\nu_e~(\bar{\nu}_e)$ and a detector resolution of $\sigma^{\mu(e)}_E~=~0.2(0.15)/\sqrt{E~{\rm{(GeV)}}}$ for muons (electrons) \cite{Acciarri:2015uup}. 
The expected number of events for the trident channels we consider in our analysis of DUNE ND are given in \tref{dunetridentevents}.

\begin{table}
\begin{center}
\begin{tabular}{|cc|cc|}
\hline
~~~~~~~~~~~$\nu$ beam &&~~~~~~~~~~~$\bar{\nu}$ beam&\\	\hline \hline
$\nu_\mu\to\nu_\mu \mu^- \mu^+$&$~~~~357$ &$\bar{\nu}_\mu\to\bar{\nu}_\mu \mu^- \mu^+$&$~~~~305$ 
\\ \hline
$\nu_e\to\nu_e \mu^- \mu^+$&$~~~~1.27$ &$\bar{\nu}_e\to\bar{\nu}_e \mu^- \mu^+$&$~~~~1.03$ 
\\ \hline 			
\end{tabular}
\end{center}
\caption{\label{tab:dunetridentevents} 
Number of expected events for trident processes in three years of operation of DUNE ND for each beam.
We do not display the numbers for processes with one or two electrons in the final state, as this will be studied in a future publication \cite{Y:Perez}, however we estimated that these have a negligible effect on the SMEFT fits.   
The $\tau$ trident events have vanishing rates. 
}
\end{table}

We can now  interpret the DUNE trident event in terms of constraints on the wEFT parameters. 
The general tree-level formula for the ratio of the trident cross section to its SM-predicted value reads 
\beq
{\sigma(\nu_b \gamma^* \to \nu_a \ell^-_c \ell^+_d)
\over \sigma_{\rm SM}(\nu_b \gamma^* \to \nu_a \ell^-_c \ell^+_d)}
=  {\sigma(\bar \nu_a \gamma^* \to \nu_b \ell^-_c \ell^+_d)
\over \sigma_{\rm SM}(\bar \nu_a \gamma^* \to \nu_b \ell^-_c \ell^+_d)}
 \approx 1 + 2 {g_{LL, \rm SM}^{abcd} \delta g_{LL}^{abcd} + g_{LR, \rm SM}^{abcd} \delta g_{LR}^{abcd} 
 \over (g_{LL, \rm SM}^{abcd})^2+(g_{LR, \rm SM}^{abcd})^2 }. 
\eeq

The SM values of the wEFT couplings $g_{LX, \rm SM}^{abcd}$ are given in \eref{WEFT_gsm}.
 Here $\delta g_{LX}^{abcd}$ parameterize the deviations from the SM prediction, and we neglect the quadratic terms in $\delta g$'s.  
The expressions  for the relevant $\delta g_{LX}^{abcd}$ in terms of the SMEFT parameters are given in  Eqs.~(\ref{eq:gll1111})-(\ref{eq:gll2222}).  
Note that, given $\delta g_{LX}^{1221} = \delta g_{LX}^{2112} = 0$,  the processes 
$\nu_\mu \gamma^* \to \nu_e \mu^- e^+$ and $\nu_e \gamma^* \to \nu_\mu \mu^+ e^-$ and their conjugates 
do {\em not} probe new physics at all (at least at tree level). 
In fact, they are just another measurement of $G_F$, which of course cannot compete with the ultra-precise determination based on  the muon decay. 
They may still be useful for calibration or normalization purposes, but by themselves they do not carry any information about BSM interactions.

The remaining trident processes do probe new physics, as indicated in the dependence of the appropriate wEFT couplings on the SMEFT coefficients. 
Here we focus on the trident process with a muon pair in the final state. 
Since the neutrino and antineutrino channels probe the same wEFT coefficient, 
we can define the following ratio: 
\bea
R_\mu & \equiv &  
{\sigma(\nu_\mu\to\nu_\mu \mu^- \mu^+) + \sigma(\bar \nu_\mu\to \bar \nu_\mu \mu^- \mu^+) 
\over 
\sigma(\nu_\mu\to\nu_\mu \mu^- \mu^+)_{\rm SM} + \sigma(\bar \nu_\mu\to \bar \nu_\mu \mu^- \mu^+)_{\rm SM}}. 
\eea 
For $R_\mu$,  the expected $\nu_e$ contribution to the $\mu^+ \mu^-$ production is much smaller than the statistical error of the measurement and can be safely neglected.  
In the following we assume the measurement error for $R_\mu$ 
will be  dominated by statistics, 
and that the central value is given by the SM prediction.  
If that is the case, the numbers in \tref{dunetridentevents} translate to the following forecast:
\beq
R_\mu  = 1 \pm 0.039,   
\eeq 
which in turn translate  into the following measurement of the wEFT coefficients 
\bea \label{eq:DUNEtrident_Rexp}
& \nnl & 
-0.039 <  2 {g_{LL, \rm SM}^{2222} \delta g_{LL}^{2222} + g_{LR, \rm SM}^{2222} \delta g_{LR}^{2222}  \over (g_{LL, \rm SM}^{2222})^2+(g_{LR, \rm SM}^{2222})^2 } < 0.039 .
\eea

\subsection{Neutrino scattering off electrons}

We turn to  neutrino scattering on electrons.
For the CC process $\nu_\mu e^- \to \mu^- \nu_e$  the threshold energy is $m_\mu^2/2m_e\sim 10.9$ GeV and therefore its rate is negligible in DUNE.\footnote{%
Even if this process were abundant in DUNE it would not probe new physics for the reasons explained below \eref{gll2222}.} 
We focus on  the NC processes 
$\nu_{\mu} e^- \to \nu_{\mu} e^-$ and $\bar \nu_{\mu} e^- \to \bar \nu_{\mu} e^-$. 
The total cross section can be expressed in terms of the wEFT parameters as 
\bea 
 \sigma_{\nu_\mu e} & = &
  {s \over 2 \pi v^4} \left [  (g_{LL}^{2211})^2  +  {1 \over 3} (g_{LR}^{2211})^2  \right ]
  \approx
    {m_e E_\nu \over \pi v^4} \left [  (g_{LL}^{2211})^2  +  {1 \over 3} (g_{LR}^{2211})^2  \right ] ,
 \nnl 
  \sigma_{\bar \nu_\mu e} & = &
  {s \over 2 \pi v^4} \left [  (g_{LR}^{2211})^2  +  {1 \over 3} (g_{LL}^{2211})^2  \right ]
  \approx   {m_e E_\nu \over \pi v^4} \left [  (g_{LR}^{2211})^2  +  {1 \over 3} (g_{LL}^{2211})^2  \right 
 ], 
\eea 
where $s=2m_eE_\nu$ is the center-of-mass energy squared of the collision, 
$E_\nu$ is the incoming neutrino energy in the lab frame, and $m_e$ is the electron mass. 
Plugging the above cross sections into \eref{nevents} and integrating over the incoming neutrino energy spectrum we obtain the total number of neutrino scattering events in the neutrino and antineutrino modes.  
The total number of the scattering events predicted by the SM is given in \tref{elastic}, where we also give the fractional contribution of $\nu_\mu$ and  $\bar \nu_\mu$ initiated processes, as well as the contribution due to  the electron neutrino contamination.   
Comparing the results of \tref{dunetridentevents} and \tref{elastic} we note that the number of events for the neutrino-electron scattering is larger than the one for trident, even though the cross sections 
are of the same order for both. 
This is a consequence of  the fact that the number of electron targets is $\mathcal{O}(10^3)$ times larger than the number of nucleus targets.

\begin{table}
\bc
\begin{tabular}{|c|c|c|c|c|c|}
 \hline 
&     $N_{\rm tot}^{\nu -e}$ & $r_{\nu_\mu}^{\nu -e}$ & $r_{\bar \nu_\mu}^{\nu -e}$ &  $r_{\nu_e}^{\nu -e}$ & $r_{\bar \nu_e}^{\nu -e}$  
\\ \hline 
$\nu$-mode & $1.69 \times 10^6$ & 0.898 &  0.059 & 0.040 & 0.003
\\ \hline 
$\bar \nu$-mode &  $1.29 \times 10^6$ & 0.103 & 0.867 & 0.013 &0.017 
\\ \hline
\end{tabular} 
\ec
\caption{ \label{tab:elastic}
Number of elastic neutrino scattering events on electrons in three years of operation of DUNE ND.  
We also give the fractional contribution to the total rate  of the four contributing processes: 
$\nu_{e/\mu} e^- \to \nu_{e/\mu} e^-$ and $\bar \nu_{e/\mu} e^- \to \bar \nu_{e/\mu} e^-$. 
}
\end{table}

We define the following observables which can be measured in DUNE: 
\beq
R_{\nu e}^i \equiv {x_i \sigma_{\nu_\mu e}  + \bar x_i \sigma_{\bar \nu_\mu e} \over
x_i \sigma_{\nu_\mu e}^{\rm SM}  + \bar x_i \sigma_{\bar \nu_\mu e}^{\rm SM} },   
\eeq  
where $x_i$ ($\bar x_i$) is the fraction of $\nu_\mu$ ($\bar \nu_\mu$) in the incoming beam in  the neutrino ($i = \nu$) and antineutrino ($i = \bar \nu$) modes.  
We assume the effect of the electron neutrino contamination of the beam can be estimated and subtracted away, and in the following analysis we approximate $x_\nu = 0.9$, $\bar x_\nu = 0.1$, and $\bar x_i = 1- x_i$. 
Writing $R_{\nu e}^i = 1 + \delta R_{\nu e}^i$, the deviation from the SM prediction can be expressed by the following combination of the wEFT parameters: 
\beq
\delta R_{\nu e}^i = 
2 {  \left ( 1 + 2 x_i \right ) \delta g_{LL}^{2211}  g_{LL, \rm SM}^{2211}  
+  \left (3 - 2 x_i \right ) \delta g_{LR}^{2211} g_{LR, \rm SM}^{2211}  \over 
 \left ( 1 + 2 x_i \right )  (g_{LL, \rm SM}^{2211})^2  
 +   \left (3 - 2 x_i \right )  (g_{LR, \rm SM}^{2211})^2   }. 
\eeq 
We assume that the error of measuring $R_{\nu  e}^i$ will be statistically dominated and that the central values will coincide with the SM prediction. 
Given the number of events in each mode displayed in \tref{elastic}, we can forecast the following constraint on the wEFT parameter 
\beq
\label{eq:DUNEelastic_Rexp}
- 8.0 \times 10^{-4} <\delta R_{\nu e}^\nu < 8.0 \times 10^{-4}, 
\qquad 
-  9.1 \times 10^{-4} <  \delta R_{\nu e}^{\bar \nu} <  9.1 \times 10^{-4}.  
\eeq

\begin{figure}[htb]
  \centering
  \includegraphics[scale=0.5]{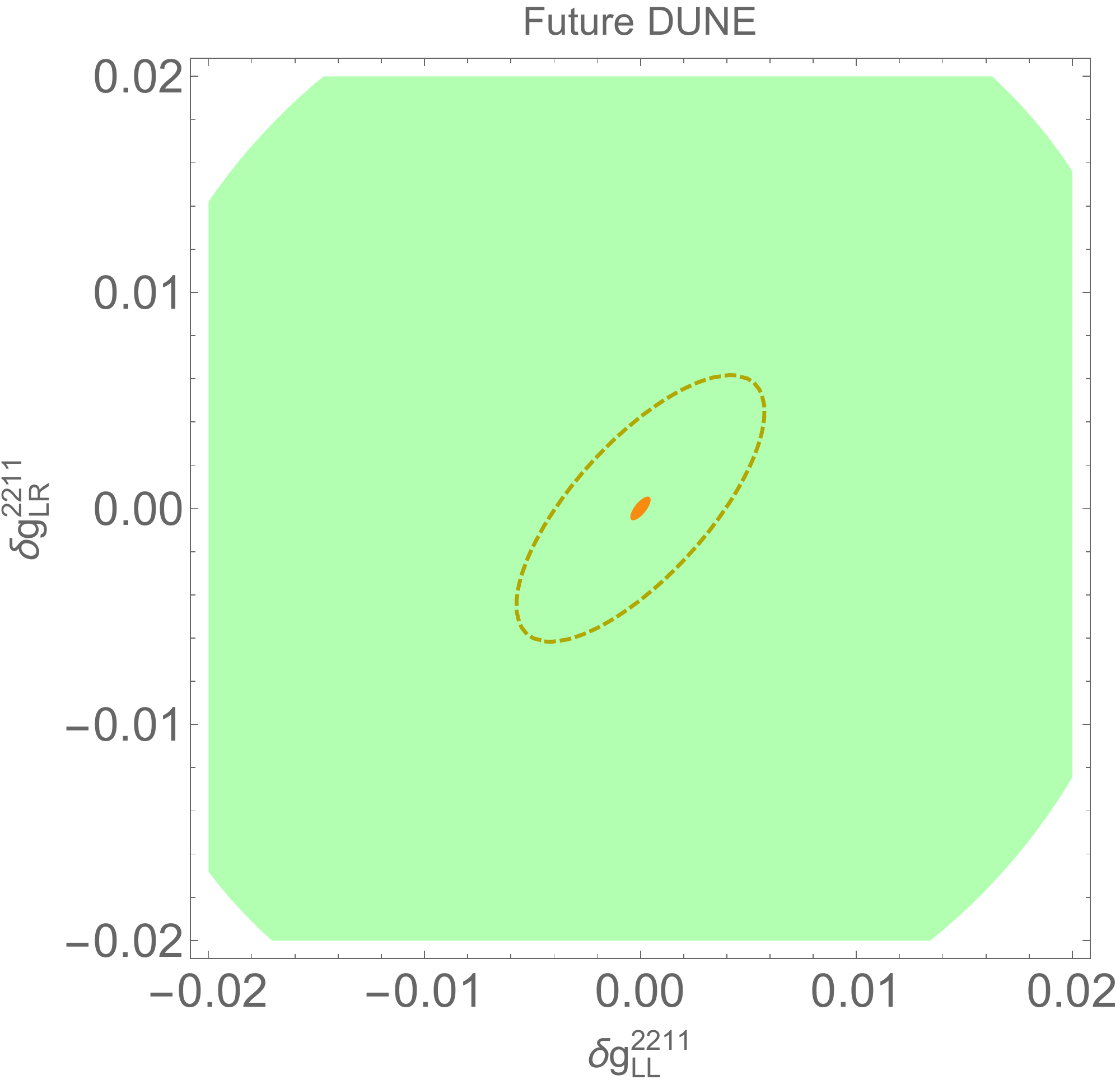}
  \caption{
Orange: projected 95\% CL constraints on the wEFT parameters  
 $\delta g^{2211}_{LL}$ and $g^{2211}_{LR}$ in \eref{WEFT_nnll} from elastic neutrino scattering on electrons in DUNE ND, assuming the measurement errors will be dominated by statistics. 
The dashed line shows the analogous constraints assuming 1\% systematic error on the $R_{\nu e}^i$ measurements.  
The green region is allowed by the past $\nu_\mu$-$e$ scattering experiments  \cite{Olive:2016xmw}. 
  }
  \label{fig:Rnue}
\end{figure}

 DUNE will by no means be the first probe of effective neutrino couplings to electrons. 
The scattering cross section of {\em muon} neutrinos and antineutrinos on electrons was measured e.g. 
 in the CHARM \cite{Dorenbosch:1988is}, CHARM-II \cite{Vilain:1994qy}, and  BNL-E734 \cite{Ahrens:1990fp} experiments with an $\cO(1)\%$ accuracy. 
The existing constraints on the wEFT parameters are combined by PDG \cite{Olive:2016xmw}:
\beq
g_{LV}^{\nu_\mu e} =  -0.040 \pm 0.015, 
\qquad g_{LA}^{\nu_\mu e} = -0.507 \pm 0.014, 
\eeq 
where in the notation of \eref{WEFT_nnll} 
$g_{LL}^{22 11 } =  (g_{LV}^{\nu_\mu e}+g_{LA}^{\nu_\mu e})/2$ and
$g_{LR}^{22 11} =  (g_{LV}^{\nu_\mu e}-g_{LA}^{\nu_\mu e})/2$. 
DUNE is expected to  dramatically improve on the existing constraints, as is evident in \fref{Rnue}, 
even under the hypothesis that  the systematic errors will greatly exceed the statistical ones.

\subsection{Neutrino scattering off nuclei}

We turn to discussing neutrino scattering off nuclei. 
To estimate the number of CC and NC scattering events expected in the SM we use the cross sections quoted in Ref.~\cite{Alion:2016uaj}. 
The results for the number of events and the fractional contributions of different neutrino flavors are given in \tref{CCandNC}.

\begin{table}
\bc
\begin{tabular}{|c|c|c|c|c|c|}
 \hline 
&     $N_{\rm tot}^{CC}$ & $r_{\nu_\mu}^{CC}$ & $r_{\bar \nu_\mu}^{CC}$ &  $r_{\nu_e}^{CC}$ & $r_{\bar \nu_e}^{CC}$  
\\ \hline 
$\nu$-mode & $4.25 \times 10^8$ & 0.964 &  0.028 & 0.007 & 0.001
\\ \hline 
$\bar \nu$-mode &  $1.74 \times 10^8$ & 0.201 & 0.790 & 0.004 &0.005 
\\ 
\hline \hline 
&     $N_{\rm tot}^{NC}$ & $r_{\nu_\mu}^{NC}$ & $r_{\bar \nu_\mu}^{NC}$ &  $r_{\nu_e}^{NC}$ & $r_{\bar \nu_e}^{NC}$  
\\ \hline 
$\nu$-mode & $1.48 \times 10^8$ & 0.956 &  0.037 & 0.006 & 0.001
\\ \hline 
$\bar \nu$-mode &  $7.58 \times 10^7$ & 0.157 & 0.835 & 0.003 &0.005 
\\ \hline
\end{tabular} 
\ec
\caption{ 
Number of charged- (CC) and neutral-current (NC) neutrino-nucleus scattering events in $3+3$ years of operation of the DUNE ND.  
The fractional contribution to the total rate of the four contributing processes are also calculated. \label{tab:CCandNC}
}
\end{table}

In order to reduce the impact of systematic uncertainties, experiments typically measure the ratio of the NC and CC neutrino or anti-neutrino scattering cross sections on nuclei, 
$\sigma^{\rm NC}/\sigma^{\rm CC}$.  
For isoscalar target nuclei,  when the incoming beam contains the fraction $x$ of neutrinos $\nu_A$ and  $\bar{x}\equiv(1-x)$ of anti-neutrinos $\bar \nu_a$ of the same flavor, the ratio of NC to CC scattering events can be written as 
\beq 
\label{eq:Rnucleons}
R_{\nu_a N} \equiv  
{x \sigma_{\nu_a N \to \nu_a N} + \bar x  \sigma_{\bar \nu_a N \to \bar \nu_a N}    
\over 
 x \sigma_{\nu_a N \to e_a^- N} + \bar x  \sigma_{\bar \nu_a N \to e_a^+ N}    }
=   
(g^{\nu_a}_{L})^2   + r^{-1} (g^{\nu_a}_{R})^2, 
\eeq 
with
\beq
r
= 
{x \sigma_{\nu_a N \to e_a^- N} + \bar x  \sigma_{\bar \nu_a N \to e_a^+ N}    
\over 
\bar   x \sigma_{\nu_a N \to e_a^- N} + x \sigma_{\bar \nu_a N \to e_a^+ N}    },  
\eeq 
and the effective couplings $g^{\nu_a}_{L/R}$ defined in \eref{gnuadef}.
This is a simple generalization of the well-known Llewellyn-Smith formula \cite{LlewellynSmith:1983tzz} usually presented in the limit $x=1$ or $x=0$.  
For any value of $x$ the dependence on the nuclear structure is contained only in the factor $r$ which can be separately measured in experiment.
In the following analysis we use \eref{Rnucleons} for a $\nu_\mu$/$\bar \nu_\mu$ incoming beam, 
with $x_\nu = 0.9$ in the $\nu$-mode, and $x_{\bar \nu} = 0.1$ in the $\bar \nu$-mode, 
in which case $r_\nu \approx 2.5$, $r_{\bar \nu} \approx 0.4$. 
In reality, for the scattering process in DUNE ND there will be important corrections to \eref{Rnucleons}.
First of all, the ${}^{40}$Ar target nuclei are not isoscalar, 
which implies corrections to the Llewellyn-Smith formula with a more complicated dependence on the nuclear structure. 
In particular, $R_{\nu_\mu N}$ will also depend on the ratio of neutrino effective couplings to up and down quarks.  
Furthermore, the incoming beam has an $\cO(1)$\% admixture of electron neutrinos.
We note that, in the general case where the neutrino effective couplings to quarks may depend on the lepton generation,  the dependence on the nuclear structure does not cancel in the ratio of NC to CC scattering events when the incoming beam contains more than one neutrino flavor. 
The impact of this DUNE measurement will crucially depend on how well these systematic effects can be controlled.  
Assuming they can be accurately constrained by dedicated measurements of CC cross sections \cite{Acciarri:2015uup},  we consider the best case scenario where the measurement errors are dominated by statistics. 
Writing $R_{\nu_\mu N}^i = R_{\nu_\mu N, \rm SM}^i (1  + \delta R_{\nu_\mu N}^i)$, 
the deviation of the ratio from the SM prediction can be constrained by DUNE as     
\begin{eqnarray}\label{eq:DUNEnu-N_Rexp}
-9.6 \times 10^{-5}<\delta R_{\nu_\mu N}^\nu<9.6 \times 10^{-5}, 
\qquad 
-1.4 \times 10^{-4}<\delta R_{\nu_\mu N}^{\bar{\nu}}<1.4 \times 10^{-4}. 
\end{eqnarray}

Expanding \eref{Rnucleons} linearly in the wEFT Wilson coefficients we get
\beq
\delta R_{\nu_\mu N}^i \simeq
2 {g^{\nu}_{L, \rm SM} \delta g^{\nu_\mu}_{L} +r_i^{-1} g^{\nu}_{R, \rm SM} \delta g^{\nu_\mu}_{R}
 \over 
(g^{\nu}_{L, \rm SM} )^2 + r_i^{-1} (g^{\nu}_{R, \rm SM})^2 
}, 
\eeq
where in terms of the parameters in \eref{WEFT_nnqq} we have 
$g^{\nu}_{X, \rm SM} \delta g^{\nu_\mu}_{X}  = 
\sum_{q=u,d} g^{\nu_\mu q}_{LX, \rm SM} \delta g^{\nu_\mu q}_{LX} - (g^{\nu}_{X, \rm SM})^2 \bar{\epsilon}_L^{\nu_\mu d} $.  

\begin{figure}[htb]
  \centering
  \includegraphics[scale=0.5]{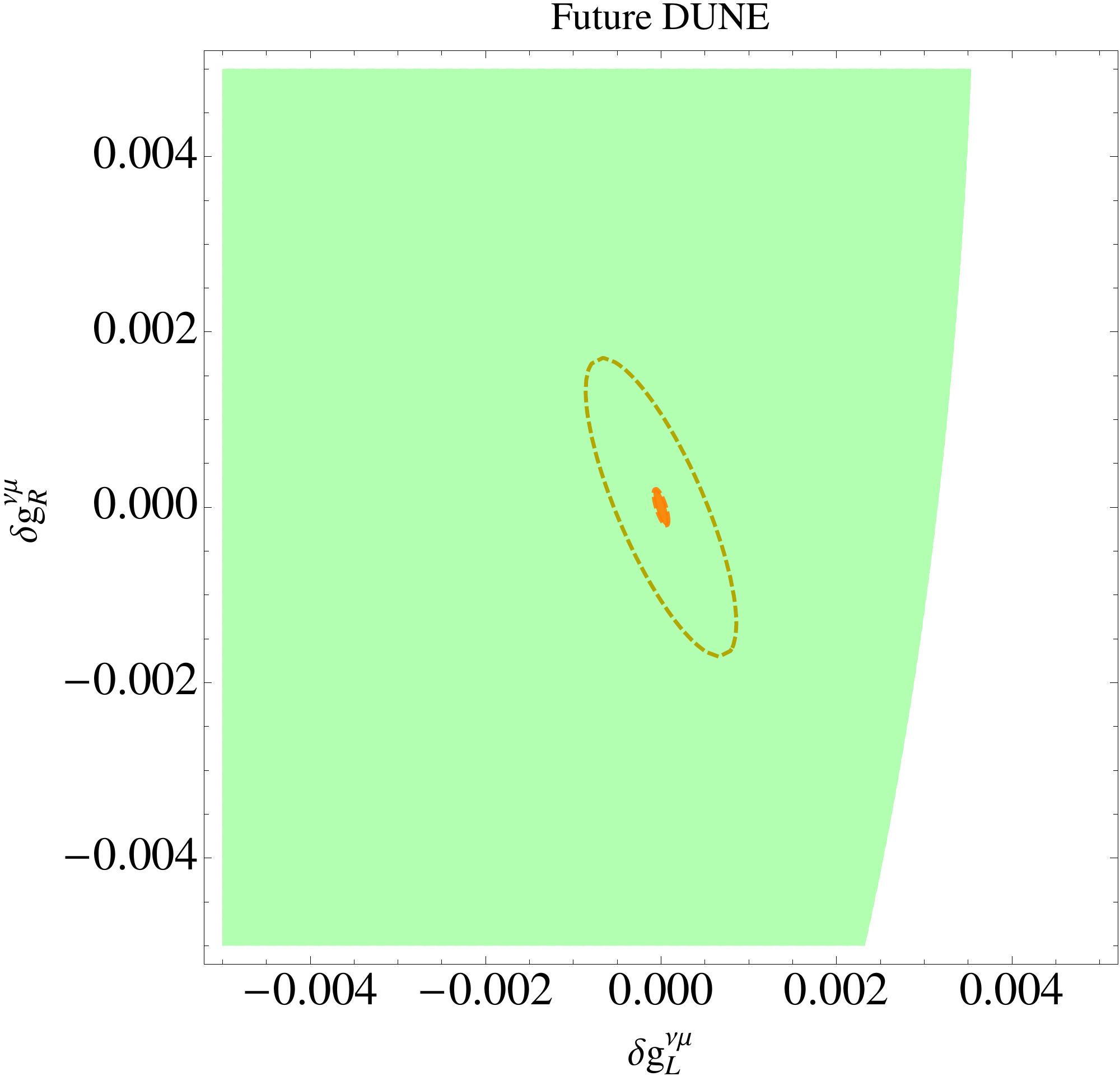}
  \caption{
Orange: projected 95\% CL constraints on the wEFT parameters  
 $\delta g^{\nu_\mu}_{L}$ and $\delta g^{\nu_\mu}_{R}$ in \eref{WEFT_nnll} from neutrino scattering off argon nuclei in DUNE ND, assuming the measurement errors will be dominated by statistics. 
The dashed line shows the analogous constraints assuming 0.1\% systematic error on the $R_{\nu_\mu N}^i$ measurements.  
The green region is allowed by the past $\nu$-$N$ scattering experiments  \cite{Olive:2016xmw}. 
  }
  \label{fig:RnuN}
\end{figure}

In the past, {\em electron} neutrino scattering on nuclei was measured in the CHARM experiment \cite{Dorenbosch:1986tb}, albeit with a poor accuracy, which is currently being improved thanks to the COHERENT experiment \cite{Akimov:2017ade}. 
Much more information is available regarding the scattering cross sections of {\em muon} neutrinos on nuclei \cite{Allaby:1987vr,Blondel:1989ev,McFarland:1997wx},  which probe the interaction terms in  \eref{WEFT_nnqq}.
The combined analysis in the PDG yields the constraints on the parameter combinations defined in \eref{gnuadef}: 
$(g_L^{\nu_\mu})^2 =   0.3005 \pm 0.0028$,  
$(g_R^{\nu_\mu})^2  =    0.0329 \pm 0.0030$, 
$\theta_L^{\nu_\mu} =  2.50  \pm  0.035$ and
$\theta_R^{\nu_\mu}    =   4.56^{+0.42}_{-0.27}$  \cite{Olive:2016xmw}. 
One can see that the couplings $g_{L/R}^{\nu_\mu}$ are probed with a relative accuracy of order $1\%$,  
which should be improved in DUNE. 
In this case the DUNE errors are unlikely to be statistics dominated, due to the nuclear structure dependent corrections to the Llewellyn-Smith formula  discussed above. 
However, even a $\cO(0.1)$\% accuracy would lead to a dramatic improvement of the existing constraints, as is shown in Fig. \ref{fig:RnuN}.

Finally, it should be noted that in a year run in the neutrino mode, the intense neutrino source will provide approximately $10^8$ total CC$+$NC neutrino interactions in a 100-t near detector; and almost 0.4 times in the antineutrino mode. Hence, thanks to the extremely high rate of neutrino interactions, one expects the near detector to be systematics dominated within the first year of the data taking. 
\vspace{1cm}
 
 In \sref{SMEFT} we will use the results in Eqs.~\eqref{eq:DUNEtrident_Rexp}, ~\eqref{eq:DUNEelastic_Rexp} and ~\eqref{eq:DUNEnu-N_Rexp} in \sref{SMEFT} to estimate the impact of the DUNE ND measurements  on the global SMEFT fit \footnote{The $0.1~(1)\%$ systematic errors mentioned in Figures 1 and 2 and Table 6 are the systematic errors in the measurement of $R$ defined in Eqs (3.3), (3.7) and (3.11). For the DUNE measurements we define 
$$\chi^2=\sum_{\nu{\&}\bar{\nu}}\delta R^2(\frac{1}{\sigma^2_{\delta R}}+\frac{1}{\sigma^2_{\rm{sys}}}),$$
with $\delta R$'s and their errors defined in Eqs (3.5), (3.8-3.9) and (3.13-3.14). The main sources of systematic uncertainties in DUNE will be on the beam flux normalization and detector performance. A careful study on the effect of different systematic uncertainties of DUNE in the measurement of the SM parameters is necessary and  will be done in a future publication \cite{Y:Perez}.
}.  
We note in passing that  effective 4-fermion interactions of neutrinos with quarks can also be probed via matter effects in neutrino oscillations \cite{Mikheev:1986gs,Gonzalez-Garcia:2013usa,Coloma:2017egw,Farzan:2017xzy}, see also \cite{Coloma:2015kiu,deGouvea:2015ndi,Masud:2015xva,Masud:2016gcl,Masud:2016bvp,Blennow:2016etl,Agarwalla:2016fkh,Deepthi:2016erc,Ghosh:2017lim,Bakhti:2016gic}
for projected DUNE constraints. Estimating the sensitivity of oscillation processes to the wEFT parameters is beyond the scope of this paper. 

\section{Related precision experiments}
\label{sec:other}

In the SMEFT framework, the neutrino couplings are correlated with the charged lepton ones due to the underlying $SU(2)_L$ symmetry.
In this context, it is important to have in mind a more global picture, and include in the analysis  other relevant experimental results which do not necessarily involve neutrinos. 

An important source of information about lepton  couplings are the measurements of the $e^+ e^- \to e_a^+ e_a^-$ differential cross sections in LEP-1 \cite{LEP:2003aa} and LEP-2 \cite{Schael:2013ita} experiments.
The LEP-1 data place per-mille level constraints on the vertex corrections $\delta g^{Zf}$, $f= q,\ell$, 
while the LEP-2 data probe $\delta g^{Wf}_L$, and the 4-fermion $LLLL$ and $LLQQ$ operators at a percent level. 
We will take into account the LEP input by directly using the likelihood function given in  Ref.~\cite{Efrati:2015eaa}.  

At lower energies,  some relevant information is provided by parity-violating M\o ller scattering,
which probes the electron's axial self-coupling in the wEFT: 
\beq
\cL_{\rm wEFT} \supset {1 \over 2 v^2} g_{AV}^{ee} \left [ - (\bar e \bar \sigma_\mu e)  (\bar e \bar \sigma_\mu e)
+  (e^c \sigma_\mu \bar e^c)  (e^c \sigma_\mu \bar e^c) \right ].
\eeq 
The SLAC E158 measurement \cite{Anthony:2005pm} can be translated into the {\em current} constraint  $g_{AV}^{ee} = 0.0190 \pm 0.0027$ \cite{Olive:2016xmw}. In the near future, the MOLLER collaboration in JLAB will significantly reduce the error on the parameter $g_{AV}^{ee}$. 
The projected accuracy on  $\sin^2 \theta_W(m_Z)$ quoted in Ref.~\cite{Benesch:2014bas}  can be recasted into the {\em future} constraint $g_{AV}^{ee}  =  0.0225 \pm 0.0006$, which will reduce the current error by almost a factor of 5. 

In the SMEFT context, there are many additional probes of the neutrino couplings to quarks, as they are correlated with charged lepton couplings.  
For example, the wEFT couplings 
\beq
\cL_{\rm wEFT} \supset - {1 \over 2 v^2}   g^{e q}_{AV} 
(\bar e\, \bar \sigma_\rho  e - e^c \sigma_\rho \bar e^c ) 
(\bar q\, \bar \sigma^\rho q + q^c \sigma^\rho \bar q^c)
\eeq
can be probed at low energies by measurements of APV \cite{Wood:1997zq} and polarization asymmetries in scattering of  electrons on nuclei \cite{Androic:2013rhu,Wang:2014bba}.   
The PDG quotes the combined constraints $g^{eu}_{AV} + 2 g^{ed}_{AV} =  0.489 \pm 0.005$ and 
$2 g^{eu}_{AV} -  g^{ed}_{AV}  =  -0.708 \pm 0.016$, 
where the SM predicts  $g^{e u, \rm SM}_{AV} =  -0.1887$ and $g^{e d, \rm SM}_{AV} =  0.3419$ \cite{Olive:2016xmw}.  
These should be significantly improved in the near future, thanks to precise determination of APV in $^{225}$Ra ions~\cite{Portela:2013twa}, 
and improved measurement of polarization asymmetries in the Qweak \cite{Androic:2013rhu}, MESA P2  targets \cite{Becker:2018ggl},  and SoLID \cite{Chen:2014psa} experiments. 
To estimate the future bounds, 
for APV and Qweak we translate the projected sensitivity to  $s_\theta^2$ \cite{Erler:2014fqa} into a constraint on  the relevant combination of  $g^{eq}_{AV}$, 
for P2 we assume the relative uncertainty on the weak charge measurement of $1.7\%$ ($0.3 \%$) for the hydrogen (carbon) target, 
while for SoLID we use the likelihood  extracted from Fig.~3 of Ref.~\cite{Zhao:2017xej}. 
Defining $\delta g^{eq}_{AV} \equiv g^{eq}_{AV} - g^{e q, \rm SM}_{AV}$, we estimate the future constraints will be per-mille level:   
$\delta g^{eu}_{AV} = (0.29 \pm 0.65)\times 10^{-3}$,  
$\delta  g^{ed}_{AV} = (-0.54. \pm 0.88)\times 10^{-3} $,  as illustrated in \fref{APV}.  

\begin{figure}[htb]
  \centering
  \includegraphics[scale=0.5]{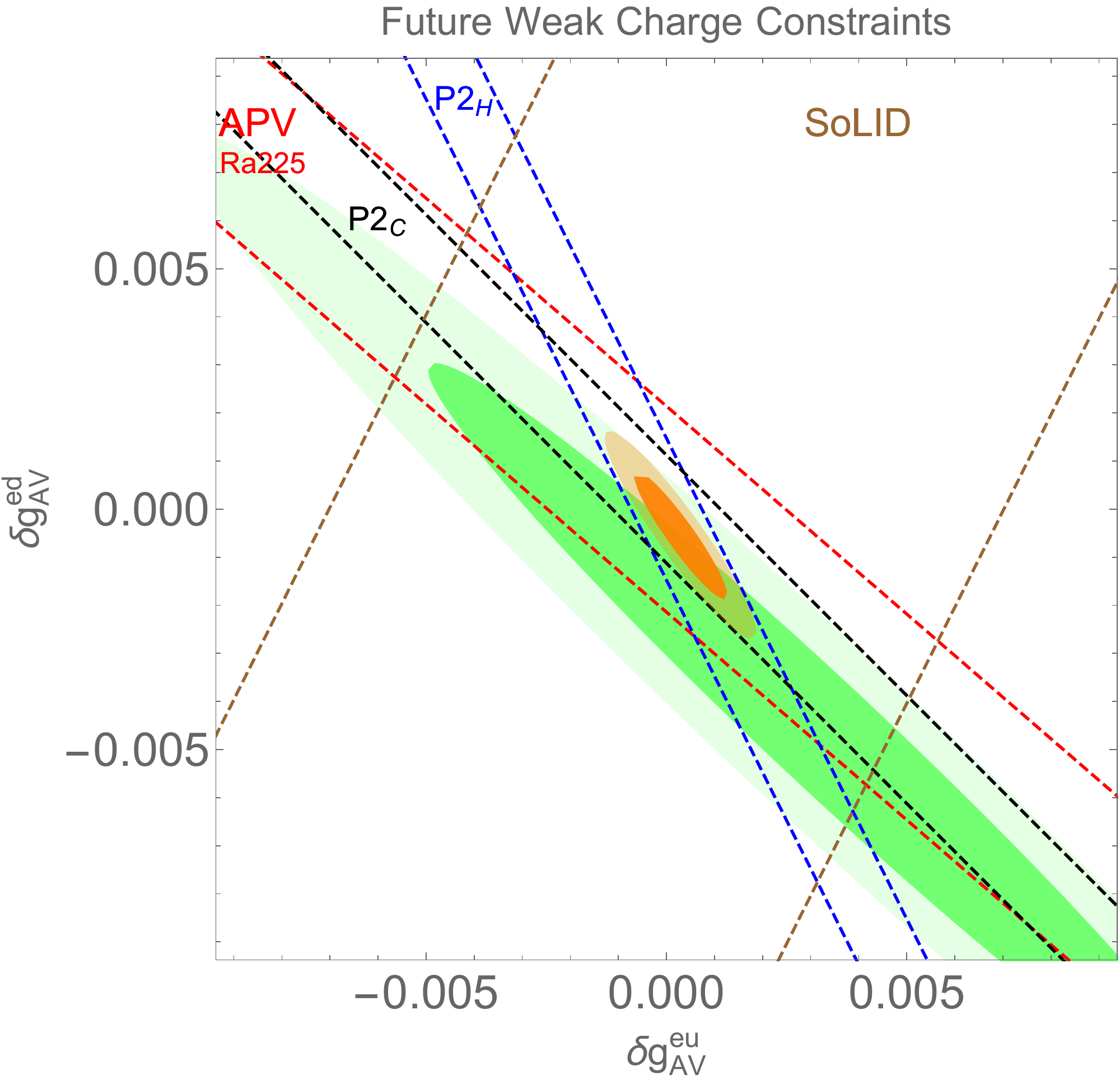}
  \caption{
 Present (green) and projected (orange) 68\% and 95 \% CL constraints on the wEFT parameters 
 $g^{e d}_{AV}$ and $g^{e d}_{AV}$. 
 The dashed lines show the projected 95 \% CL constraints separately from  P2 with the hydrogen (blue) and carbon (black) target, SoLID (brown), and APV in $^{225}$Ra (red).  
  }
  \label{fig:APV}
\end{figure}

At higher energies, where the wEFT is no longer valid and the full SMEFT formalism needs to be used,  lepton-quark interactions can be probed in colliders.   
Existing LEP measurements \cite{LEP:2003aa,Schael:2013ita} of electron-quark interactions are compiled in \cite{Falkowski:2017pss} with the conclusion that, in combination with the low-energy measurements, chirality-conserving 2-electron-2-quark operators are constrained in a model-independent way at a percent level.  
The sensitivity improves greatly when lepton pair Drell-Yan production at the LHC is taken into account, and in fact 2-lepton-2-quark operators can be accessed for all three lepton generations \cite{Cirigliano:2012ab,deBlas:2013qqa,Greljo:2017vvb}. 
For $ee qq$ and $ee dd$ operators the current sensitivity is at a per-mille level, and is expected to improve to an $\cO(10^{-4})$ level after the high-luminosity run is completed  \cite{Greljo:2017vvb}.
That level of precision may be difficult to beat in DUNE.  
However,  a full model-independent analysis of the LHC constraints has not been attempted yet, due to a large number of different  SMEFT operators contributing to the Drell-Yan process.
In this analysis we focus on low-energy precision measurements, and we do not include the LHC constraints. 
It is reasonable to expect that the LHC will leave flat or poorly constrained directions in this parameter space, along which  Wilson coefficient much larger than $\sim 10^{-4}$  can be present without  conflicting the data. 
The task of lifting these flat directions would then be left for  DUNE and other low-energy precision experiments.

\section{Future evolution of SMEFT constraints}
\label{sec:SMEFT}

In this section we translate the projected low-energy constraints discussed in Sections~\ref{sec:dune}~and~\ref{sec:other} into the SMEFT language.
The goal is to illustrate the place of DUNE in the landscape of low-energy precision observables, 
and to quantify its sensitivity to new physics at high-energy scales.  
With this goal in mind, we consider deformations of the SM  where only a single SMEFT parameter is non-zero at a time. 
Then we compare the constraints on that parameter using the current and future low-energy experiments. 
For the current ones we use  the likelihood function compiled in Ref.~\cite{Falkowski:2017pss}.  
This is subsequently combined with the future DUNE constraints based on the projections in \sref{dune}, and with the future APV and polarization asymmetry constraints discussed in \sref{other}. 
To illustrate DUNE's impact on the precision program we separately show the future projections with and without the DUNE input. 
Furthermore, the DUNE projections are shown for 3 scenarios:
(very) {\em optimistic} with the error dominated by statistics,    
{\em more realistic} with $10^{-3}$ systematic errors, and {\em pessimistic} with  $10^{-2}$ systematic errors.
\begin{table}
\begin{tabular}{|c|c|c|c|c|c|}
\hline
Coefficient & $\Delta$(current) & $\Delta$ (no sys.) & $\Delta$ ($0.1\%$ sys.) &  $\Delta$ ($1\%$ sys.) &   $\Delta$ (w/o DUNE) \\  \hline \hline
 $\delta g^{We}_L$ & 3.5 & 0.37 & 2.5 & 3.4 & 3.5 \\
$\delta g^{Z\mu}_L$ & 3.7 & 0.18 & 1.1 & 3.5 & 3.7 \\
$\delta g^{Zu}_L$  & 1.9 & 0.34 & 1.4 & 1.5 & 1.5  \\
$\delta g^{Zu}_R$  & 9.5 & 0.57 & 2.0 & 2.3 & 2.3 \\
$\delta g^{Zd}_L$  & 1.9 & 0.28 & 1.4 & 1.6 & 1.6 \\
$\delta g^{Zd}_R$  & 9.7 & 1.1 & 3.0 & 3.1 & 3.1  \\
$\delta g^{W q_1}_R$ &  1.9 & 0.36 & 1.7  & 1.9 & 1.9 \\
$[c_{\ell \ell}]_{1122}$ & 28 & 2.6 & 2.6 &21 & 28 \\
$[c_{\ell e}]_{2211}$  & 45 & 3.1 & 3.1 & 27 & 45 \\
$[c_{\ell \ell}]_{2222}$ & 2100 & 310  & 310 & 310 & 2100 \\
$[c_{\ell e}]_{2222}$ & 6300  & 970 & 970 & 970 & 6300 \\
$[c_{\ell q}^{(3)}]_{1111}$  & 1.9  & 0.36 & 1.7 & 1.9 & 1.9 \\
$[c_{\ell q}^{(3)}]_{2211}$ & 12 & 1.8 & 10 & 12 & 12  \\
$[c_{\ell q}]_{2211}$ & 210 & 3.0 & 30 & 180 & 210 \\
$[c_{\ell u}]_{2211}$ & 190 & 1.2 & 9.5 & 85 & 190 \\
$[c_{\ell d}]_{2211}$& 370 & 2.4 & 19 & 170 & 370 \\ \hline 
\end{tabular}
\caption{
\label{tab:dunesmeft}
{\bf  $1 \sigma$ uncertainty $\Delta$ in units of $10^{-4}$} on SMEFT Wilson coefficient from current and future low-energy precision measurements,  assuming only one Wilson coefficient is non-zero at a time. 
We display only those coefficients in the Higgs basis for which the constraints can be improved by at least a factor of five in the best-case scenario. 
The current uncertainties are extracted from Ref.~\cite{Falkowski:2017pss}, while the future ones also include projections from $\nu$ scattering on electrons and nucleons and trident production in DUNE, as well from APV and parity-violating electron scattering experiments discussed in \sref{other}. 
The future constraints are shown under three different hypotheses about the size of systematic errors in DUNE. 
The final column shows the future projections without the DUNE input. 
}
\end{table}
\begin{figure}[htb]
  \centering
  \includegraphics[scale=0.5]{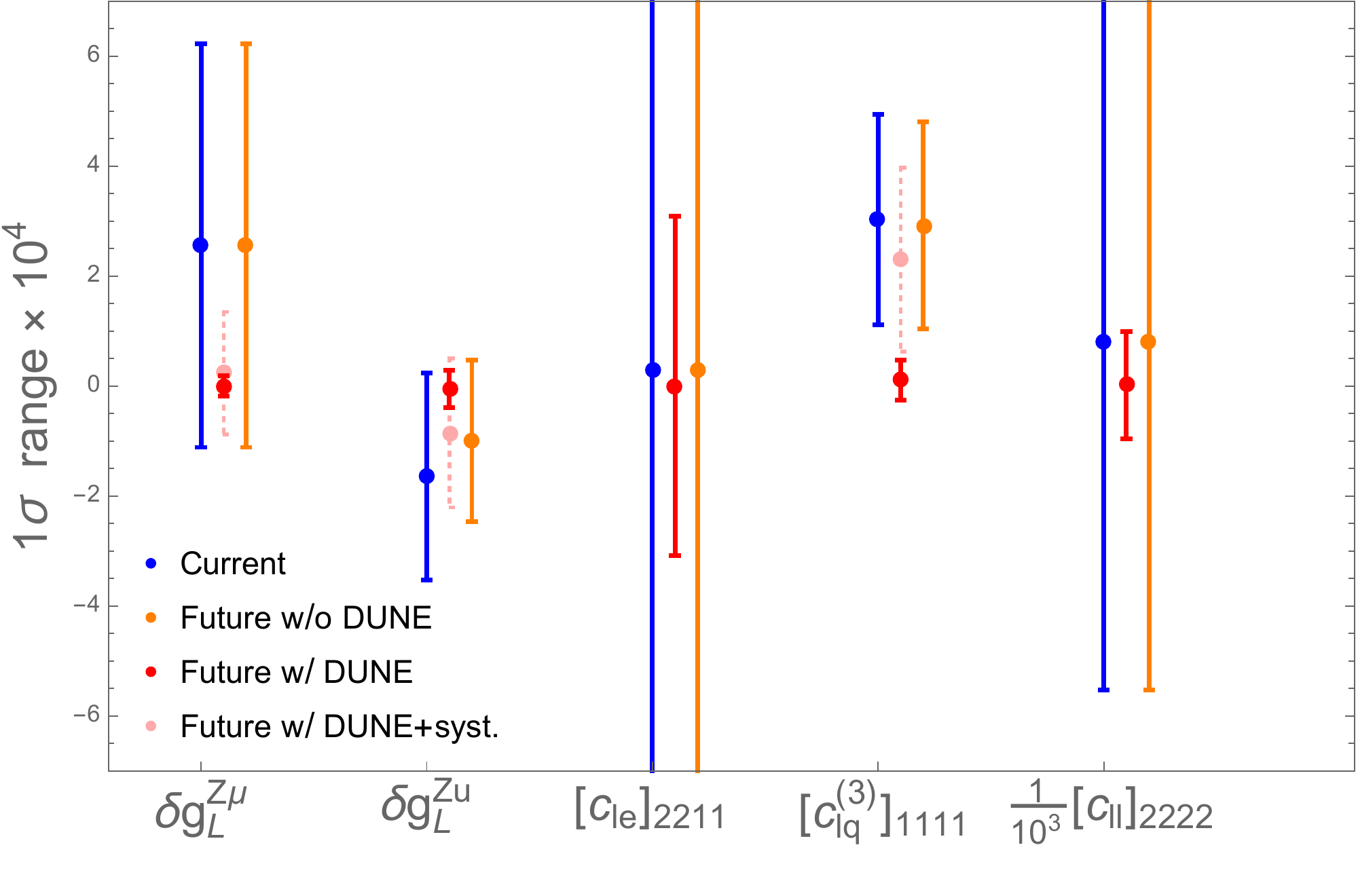}
  \caption{   \label{fig:ressmeft}
$1 \sigma$ range for selected SMEFT Wilson coefficient from current and future low-energy precision measurements,  assuming only one Wilson coefficient in the Higgs basis is non-zero at a time. 
The current constraints are represented by blue error bars.  
We future constraints are shown with the DUNE input assuming only statistical errors (red)  or $0.1\%$ systematic errors (dotted pink), and also without the DUNE input (orange). 
  }
\end{figure}

Our results are summarized in \tref{dunesmeft} where we display the constraints on those Higgs basis parameters for which the projected error is reduced by at least a factor of 5 compared to the current one.
We can see that DUNE will potentially have a dramatic impact on constraining the SMEFT parameter space. 
Without the DUNE input, the precision that can be achieved is typically an order of magnitude worse 
(for the operators we consider and given the future experiments  we take into account in our analysis).
An $\cO(10^{-4})$ relative precision can be achieved for some parameters such as the $Z$ boson coupling to muons or light quarks or certain 4-fermion $LLQQ$ operators. 
This translates into $\cO(30)$ ($\cO(300)$) TeV indirect reach for new BSM particles, assuming they are weakly (strongly) coupled to the SM muons and light quarks.  
Those scales are well beyond the direct reach of current and near-future colliders, and also surpass the indirect reach of the past electroweak precision tests in the LEP $e^+ e^-$ collider.   
Even for the more realistic scenario about systematic errors  the improvement in the reach for new physics is spectacular in some cases.   
On the other hand, in the pessimistic scenario the impact of DUNE on the precision program is marginal, except in the 4-muon sector where the existing loose trident constraints can be improved by an order of magnitude.   
These considerations highlight  the importance of precision measurements  in DUNE and the efforts to reduce experimental and theoretical sources of systematic errors.

Technically there is no problem including the future projections  into a fully global  SMEFT analysis akin to that in Ref.~\cite{Falkowski:2017pss} where all dimension-6 operators are present simultaneously. 
Such a step will be crucial once the real data are available, as only a global analysis is basis independent and  contains the complete  information about the constraints.  
However, in the present case such an exercise is not very illuminating and we do not show  global fit projections here. 
The reason is that the DUNE and other observable we consider in this paper constrain only a limited number of linear combinations of the Wilson coefficients, leaving many weakly constrained directions in the SMEFT parameter space.  
As a result, the improvement in sensitivity is hardly visible in our global fit projection, 
again with the exception of the 4-muon operators.

\section{Conclusions}
\label{sec:conclusions}

In this work, we investigated the precision reach in the determination of the SMEFT Wilson coefficients relevant for low-energy experiments. 
We studied observables related to trident production, neutrino scattering off electrons and neutrino scattering off nuclei at DUNE ND, while leaving for a future work an estimate of the sensitivity of oscillation processes to the SMEFT parameters. 
Moreover, information from parity violating M\o ller scattering, APV,  and polarization asymmetry experiments is also included, as they are sensitive to the same SMEFT parameters as  the ones probed by DUNE.  
Our projections are combined and compared with the current constraints on the SMEFT parameters. 

Our main results are summarized in \tref{dunesmeft} and illustrated in \fref{ressmeft}. 
There we assume the presence of only one non-zero Wilson coefficient at a time. 
Working in the Higgs basis of the SMEFT, the constraints on seven vertex corrections and on nine four-fermion operators are improved by at least a factor of five in the best case scenario. 
With the optimistic assumption that the DUNE errors will be dominated by statistics, 
one can reach an $\mathcal{O}(10^{-4})$ relative precision for $Z$ boson couplings to muons and light quarks, and for some 4-fermion $LLQQ$ operators. 
This could probe the new physics scale $\Lambda > \mathcal{O}(30) (\mathcal{O}(300))$ TeV, for weakly (strongly) coupled new physics to the SM muons or light quarks.
This is  beyond the direct reach of the  LHC or near-future colliders, as well as beyond the indirect reach  of electroweak precision measurements at LEP. 
Without the DUNE input, the expected precision is typically an order of magnitude worse. 
The projections are degraded with less optimistic assumptions about the systematic errors achievable in DUNE, which encourages future efforts to reduce experimental and theoretical sources of these errors.

\acknowledgments GGdC and ZT are supported by Funda\c{c}\~ao de Amparo
\`a Pesquisa do Estado de S\~ao Paulo (FAPESP) under contracts 16/17041-0 and 16/02636-8. A.F is partially supported by the European Union's Horizon 2020
research and innovation programme under the Marie Sklodowska-Curie
grant agreements No 690575 and  No 674896. ZT is particularly grateful to the hospitality of LPT Orsay where this work was initiated and acknowledges useful discussions with Yuber F. Perez. We also thank Paride Paradisi for pointing out a typo in an earlier version of Eq. 3.11.

\bibliography{paperEFT}
\bibliographystyle{JHEP}

\end{document}